\documentclass[aps,reprint,superscriptaddress,amsmath,amssymb,longbibliography]{revtex4-2}

\usepackage{graphicx}
\usepackage{xcolor}
\usepackage{amsmath}
\usepackage{bm}
\usepackage{float}
\usepackage{hyperref}
\usepackage{url}
\usepackage{makecell}
\usepackage{multirow}
\usepackage{dcolumn}
\usepackage{array}

\newcolumntype{M}[1]{>{\centering\arraybackslash}m{#1}}
\newcommand{\leftcol}{0.8in}
\newcommand{\midcola}{2.8in}
\newcommand{\midcolb}{0.7in}
\newcommand{\midcolc}{0.7in}
\newcommand{\rightcol}{1.2in}

\begin{document}

\title{Ising dynamics on multilayer networks with heterogeneous layers}
\author{Suman S. Kulkarni}
\email{sumank@sas.upenn.edu}
\affiliation{Department of Physics and Astronomy, University of Pennsylvania, Philadelphia, PA, 19104, USA}
\author{Christopher W. Lynn}
\affiliation{Department of Physics, Yale University, New Haven, CT, 06511, USA}
\affiliation{Quantitative Biology Institute, Yale University, New Haven, CT, 06511, USA}
\affiliation{Wu Tsai Institute, Yale University, New Haven, CT, 06510, USA}
\author{Mason A. Porter}
\affiliation{Department of Mathematics, University of California,
Los Angeles, CA, 90095, USA}
\affiliation{Department of Sociology, University of California,
Los Angeles, CA, 90095, USA}
\affiliation{Santa Fe Institute, Santa Fe, NM, 87501, USA}
\author{Dani S. Bassett}
\affiliation{Department of Physics and Astronomy, University of Pennsylvania, Philadelphia, PA, 19104, USA}
\affiliation{Santa Fe Institute, Santa Fe, NM, 87501, USA}
\affiliation{Department of Bioengineering, University of Pennsylvania, Philadelphia, PA, 19104 USA}
\affiliation{Department of Electrical \& Systems Engineering,
University of Pennsylvania, Philadelphia, PA, 19104 USA}
\affiliation{Department of Neurology, University of Pennsylvania, Philadelphia, PA, 19104 USA}
\affiliation{Department of Psychiatry, University of Pennsylvania, Philadelphia, PA, 19104 USA}
\affiliation{Montreal Neurological Institute, McGill University, Montreal, QC H3A 2B4, Canada}

\date{\today}


\begin{abstract}

Multilayer networks provide a framework to study complex systems with multiple types of interactions, multiple dynamical processes, and/or multiple subsystems. When studying a dynamical process on a multilayer network, it is important to consider how both layer structure and heterogeneity across layers impacts the overall dynamics. As a concrete example, we study Ising dynamics on multilayer networks and investigate how network structure affects its qualitative features. We focus primarily on multiplex networks, which are multilayer networks in which interlayer edges occur only between manifestations of the same entity on different layers, although we also consider one empirical example with a more general multilayer structure. We use numerical simulations and a mean-field approximation to examine the steady-state behavior of the Ising dynamics as a function of temperature (which is a key model parameter) for a variety of two-layer multilayer networks from both models and empirical data. We examine both the steady-state behavior and a metastable state in which the two layers are anti-aligned, and we explore the effects of interlayer coupling strength and structural heterogeneity. In synthetic multilayer networks with core--periphery structure, we show that interlayer edges that involve peripheral nodes can exert more influence than interlayer edges that involve only core nodes. Finally, we consider empirical multilayer networks from biological and social systems. Our work illustrates how heterogeneity across the layers of a multilayer network influences dynamics on the whole network. 
\end{abstract}

\maketitle


\section{Introduction}

Dynamical processes on networks are useful models of a wide variety of phenomena, including the spread of infectious diseases and other contagions, opinion dynamics, the collective behavior of coupled oscillators, and many other processes across physical, biological, and social systems \cite{barrat2008dynamical, porter2016dynamical}. The behavior of dynamical processes on networks can be sensitive to the underlying network structure. For example, network structure influences the critical threshold for disease outbreaks \cite{pastor2015epidemic}, the likelihood of consensus or polarization in opinion models \cite{noorazar2020classical,noorazar2020recent,starnini2025}, and the onset and extent of synchronization of coupled oscillators \cite{rodrigues2016kuramoto,arenas2008synchronization}. 

Many real-world systems consist of entities that interact in complex patterns. Such interactions can encompass relationships of different types, multiple subsystems, interactions at different spatial scales, and interactions that evolve with time \cite{newman2018networks,vision2020}. Multilayer networks provide a framework to encode the structure of such networked systems \cite{kivela2014multilayer, bianconi2018multilayer,aleta2019multilayer,de2022}. In most situations, dynamical processes on multilayer networks do not reduce to equivalent processes on a single network that one obtains by aggregating the layers of a multilayer network \cite{de2023more}. Moreover, because network structure can have significant impacts on the qualitative and quantitative behavior of dynamical processes, it is important to examine how the structure of individual layers and the interactions between layers affect processes on multilayer networks as a whole \cite{de2016physics}. In many systems, different layers of a multilayer network have distinct structural features \cite{kivela2014multilayer,de2022}. For example, social systems include multiple types of relationships---such as familial ties, professional relationships, and online connections---and different types of relationships have distinct connectivity patterns \cite{atkisson2020understanding}. In neural systems, neurons can communicate via synaptic connections or through extrasynaptic volume transmission, and the networks of different communication modes have distinct structural properties \cite{bentley2016multilayer,ripoll2023neuropeptidergic}. Accordingly, it is important to analyze dynamical processes on multilayer networks with structurally heterogeneous layers. To develop intuition for how such heterogeneity affects dynamical processes, we study an Ising model on multilayer networks as a concrete example. 

The Ising model is a prototypical and well-studied model in statistical physics \cite{kuelske2025ising,goldenfeld2018lectures}. It was developed originally to model ferromagnetism, but it has now been employed in studies of numerous and diverse other phenomena \cite{sethna2021statistical}. For example, it has been used as a toy model for opinion dynamics and other social dynamics~\cite{castellano2009statistical,devauchelle2024dislike,galesic2019statistical,olsson2024analogies,mullick2025}. In neuroscience, pairwise maximum-entropy models (which have been used to describe neuronal activity) are equivalent to an Ising model \cite{schneidman2006weak}. The behavior of the Ising model has been studied extensively on lattices \cite{kardar2007statistical,goldenfeld2018lectures}, networks with certain specific degree distributions \cite{dorogovtsev2002ising,herrero2004ising,bianconi2002mean}, and networks with mesoscale features \cite{chen2018double}. Such investigations have provided insights into phase transitions and critical phenomena on networks.

Previous studies have examined Ising models on multiplex networks \cite{krawiecki2018ferromagnetic} and on other networks, such as coupled networks \cite{bolfe2018ising_coupled,suchecki2009bistable, suchecki2006ising} and modular networks \cite{dasgupta2009phase}, that one can view as multilayer networks. Another relevant study is~\cite{jang2015ashkin}, which considered a mean-field approximation of the Ashkin--Teller model---which one can view as a variant of the Ising model---on a two-layer network in which each layer consists of the same monolayer network. We particularly highlight Ref.~\cite{krawiecki2018ferromagnetic}, which examined an Ising model on two-layer multiplex networks. In this model, each node shares the same spin across the two layers. By contrast, in the present paper, we consider an Ising model in which each layer has distinct spins. Additionally, prior studies of Ising models on coupled networks \cite{bolfe2018ising_coupled,suchecki2009bistable,dasgupta2009phase} considered networks with the same architecture, whereas we explicitly examine multilayer networks with structural heterogeneity across layers. We aim to examine how the structure of each layer and the coupling between layers affect the overall qualitative features of Ising dynamics on multilayer networks. To do this, we use both Monte Carlo (MC) simulations and a mean-field approximation and consider both synthetic and empirical networks. We primarily study multiplex networks, where interlayer edges can exist only between entities and their counterparts in different layers, although one of our empirical examples consists of coupled Facebook networks and has interlayer edges between different entities in different layers.

By studying two-layer multiplex networks with interlayer edges of uniform weight, we investigate Ising dynamics as a function of interlayer coupling strength (i.e., the interlayer edge weight) and other model parameters. We first discuss the steady-state behavior of the ferromagnetic Ising model on a multiplex network with Erd\H{o}s--R\'{e}nyi (ER) layers with identical connection probabilities. The model has an order--disorder transition, and increasing the interlayer coupling increases the transition temperature. Using a mean-field approximation, we examine a metastable solution in which the two layers have magnetizations of opposite signs (i.e., this is an ``anti-aligned" configuration) and we illustrate that increasing the stronger interlayer coupling shrinks the temperature range that supports this metastable state (i.e., ``phase"). We then explore Ising dynamics on multilayer networks with layers with different mean degrees by considering ER layers with different connection probabilities. Unsurprisingly, in this situation, the steady-state configuration has a parameter regime in which the denser layer is more ordered than the sparser layer. Moreover, the sparser layer retains a nonzero magnetization past its original critical temperature due to its coupling to the denser layer. The heterogeneity in the mean degrees also decreases the range of interlayer coupling values $r$ that supports the metastable anti-aligned configuration.
  
To examine Ising dynamics on multilayer networks with heterogeneous mesoscale structures, we consider networks with a core--periphery (CP) structure on one layer and an ER structure on the other layer. Counterintuitively, we find that interlayer connections that involve peripheral nodes can exert more influence on the system's dynamics than interlayer connections that involve only core nodes. Finally, we examine Ising dynamics on empirical two-layer networks. We consider a network of neuronal signaling in the nematode \emph{Caenorhabditis elegans} and a multi-university Facebook network. Our research contributes to the understanding of how the structure of individual layers in a multilayer network and the coupling between layers (see, e.g., \cite{radicchi2013abrupt,buldu2018}) can shape the overall qualitative behavior of a dynamical process on a multilayer network \cite{de2016physics,de2023more}.

Our paper proceeds follows. In Section \ref{sec:formulation}, we formulate the Ising model that we consider in our paper. In Sections \ref{sec:ER_graphs} and \ref{sec:CP_graphs}, we examine this Ising model on synthetic multiplex networks. In Section \ref{sec:real_nets}, we study this Ising model on two real-world networks. We conclude in Section \ref{sec:discussion} with a summary, a discussion of our results, and some thoughts on future research directions. Our code and a subset of our data are available at \href{https://github.com/SumanSKulkarni/Ising-Multilayer}{https://github.com/SumanSKulkarni/Ising-Multilayer}.

\section{Model formulation}\label{sec:formulation}


\subsection{Multilayer networks}

Multilayer networks provide a framework to systematically encode the connectivity of networked systems with multiple types of interactions, multiple subsystems, and other features \cite{kivela2014multilayer, aleta2019multilayer}. A multilayer network has layers in addition to nodes and edges. The layers can represent different types of relationships, different subsystems, different time points, interactions at different scales, or other things. A node in a multilayer network represents an agent or some other entity, and a node can exist in one or more layers. We use the term \emph{physical node} to refer to a node regardless of its layer and the term \emph{state node} to refer to an instantiation of a node in a specific layer. The edges within each layer are called \emph{intralayer} edges, and edges that connect nodes across distinct layers are called \emph{interlayer} edges. An important special type of multilayer network is a \emph{multiplex network}. In a multiplex network, interlayer edges only connect state nodes that correspond to the same physical node. That is, they connect different instantiations of the same entity. Multiplex networks are useful for studying multirelational systems~\cite{kivela2014multilayer,de2022}, and they provide a convenient way to encode different types of relationships between the same set of physical nodes. Interlayer edges capture how state nodes influence each other across layers.

\begin{figure}
    \centering
    \includegraphics[scale=0.8]{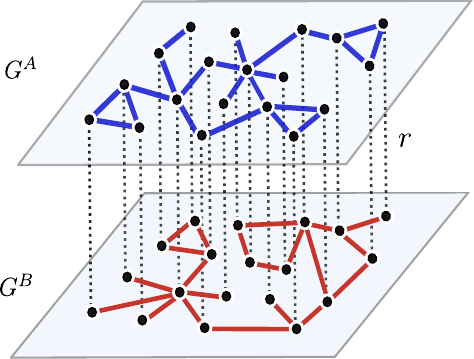}
    \caption{An example of a two-layer multiplex network in which the layers have distinct network structures. Each layer consists of a distinct graph, which we denote by $G^{A}$ and $G^{B}$, with a distinct network structure. The dotted lines represent interlayer edges that connect instances of the same node across the two layers, and the solid lines represent intralayer edges. In this example, we assume that all interlayer edges have the same weight $r$. }
     \label{fig:multiplex_schematic}
\end{figure}

In our paper, we primarily examine multiplex networks with different connectivity patterns in each layer. Consider a two-layer multiplex network with $N$ physical nodes and layers $A$ and $B$. We consider only undirected networks. A multilayer network has an associated supra-adjacency matrix $W$ to represent the network \cite{kivela2014multilayer}. For a two-layer multiplex network, the supra-adjacency matrix takes the form
\begin{equation}
    W = \begin{pmatrix}
        W^{A} & K \\
        K & W^{B}
    \end{pmatrix}\,,
\end{equation}
where $W^{A}$ and $W^{B}$ are the adjacency matrices of the networks in layers $A$ and $B$, respectively, and $K$ encodes the interlayer edges that connect the two layers. In general, interlayer edges can have different weights, but we assume for simplicity that all interlayer edges have the same weight (i.e., coupling strength) $r$. This enables us to systematically study Ising dynamics on multilayer networks as a function of the interlayer coupling strength. With this assumption, the supra-adjacency matrix is
\begin{equation}
    W = \begin{pmatrix}
        W^{A} & rI \\
        r I & W^{B}
    \end{pmatrix}\,,
\end{equation}
where $I$ is the $N \times N$ identity matrix. In Figure \ref{fig:multiplex_schematic}, we show a schematic illustration of a two-layer multiplex network with heterogeneous layers and interlayer edges of uniform strength $r$.


\subsection{Ising dynamics on ordinary networks}


An Ising model consists of discrete variables, called ``spins", which can take the values $+1$ or $-1$ \cite{sethna2021statistical}. Each node of a network (i.e., monolayer network) has a spin and interacts with its neighboring nodes. In the ferromagnetic Ising model, interactions between nodes favor alignment. A ``spin configuration" consists of the set of values ($+1$ or $-1$) of all nodes in a system.  An Ising system seeks spin configurations with low energy, but thermal fluctuations, which we quantify by the system ``temperature" $T$, disrupts this tendency. 

In the absence of an external field, the energy (i.e., Hamiltonian) of a spin configuration $\mathbf{s}$ with entries $s_i$ is
\begin{equation}
    \mathcal{H}(\mathbf{s}) = - \sum_{\langle i, j \rangle} J_{ij} s_i s_j\,,
    \label{eq:hamiltonian}
\end{equation}
where $J_{ij}$ is the strength of the interaction between node $i$ and node $j$. In the ferromagnetic Ising model, which favors alignment between neighboring nodes, $J_{ij} > 0$. The probability of a spin configuration is given by the Boltzmann probability distribution
\begin{equation}
    P_{T}(\textbf{s}) = \frac{e^{- \mathcal{H}(\textbf{s})/T}}{\sum_{\textbf{s}}e^{- \mathcal{H}(\textbf{s})/T}} = \frac{e^{- \mathcal{H}(\textbf{s})/T}}{\mathcal{Z}}\,,
\end{equation}
from which one computes the expected value of a quantity $f$ by calculating
\begin{align}
    \langle f \rangle &= \sum_{\mathbf{s}} f(\mathbf{s}) P_{T}(\mathbf{s})\,.
\end{align}

At low temperatures, the node spins tend to align with each other. As the temperature increases, thermal fluctuations increasingly disrupt this tendency. Depending on network structure, there may be a phase transition between an ordered (i.e., ferromagnetic) state \footnote{Some researchers use the term ``phase'' instead of state.} and a disordered (i.e., paramagnetic) state \cite{sethna2021statistical} at some critical temperature $T_c$. For example, the Ising model on a one-dimensional (1D) lattice does not have a phase transition in the thermodynamic limit (i.e., in the limit of infinite system size), as thermal fluctuations dominate for all temperatures $T > 0$ \cite{kardar2007statistical}. By contrast, the Ising model on a two-dimensional (2D) square lattice and three-dimensional (3D) cubic lattice both have phase transitions in the thermodynamic limit \cite{kardar2007statistical}. For configuration-model networks with power-law degree distributions, the presence of a phase transition in the thermodynamic limit depends on the power-law exponent~\cite{herrero2004ising}.


\subsection{Ising dynamics on multiplex networks}

We now define Ising dynamics on multiplex networks. We let each state node of a multiplex network have a distinct Ising spin. In general, the interlayer edges in a multilayer can have different weights. However, for simplicity, we consider two-layer multiplex networks in which all interlayer edges have the same weight $r$. This allows us to systematically study Ising dynamics as a function of the temperature $T$ and the interlayer coupling strength $r$ for any two intralayer-network structures. 

To quantitatively examine Ising dynamics on a multiplex network, we monitor both overall magnetization and single-layer magnetization. The overall magnetization is
\begin{equation}\label{eq:m_overall}
    m = \frac{\sum_{i}^{S} s_i}{S}\,,
\end{equation}
where $S = 2N$ is the total number of state nodes (and hence spins) of a network and $N$ is the total number of physical nodes. We also track the magnetization of each individual layer. The magnetizations of layers $A$ and $B$ are  
\begin{equation}\label{eq:m_AB}
    m_{A} = \frac{\sum_{i}^{N} s_i}{N}\,, \quad m_{B} = \frac{\sum_{N + 1}^{S} s_i}{N}\,. \vspace{1mm}
\end{equation}
In our numerical simulations, we use the absolute value of the magnetization in practice because the Hamiltonian \eqref{eq:hamiltonian} is invariant if we flip all of the spins.

In networks with finitely many state nodes, MC simulations can flip the sign of the magnetization of a system through a series of local spin flips \cite{sandvik-mc}. This can yield a mean magnetization that is close to $0$ when we average over a large number of MC iterations, even in the ordered state. To avoid this issue, we monitor the mean absolute values of the overall magnetization and the layer magnetizations.

\begin{table*}[htb]
\centering
\caption{\label{tab:networks} The two-layer multiplex networks on which we examine Ising dynamics.}
\renewcommand{\arraystretch}{1.2} 
\setlength{\tabcolsep}{2pt}
\addtolength{\tabcolsep}{2pt}
\begin{tabular}{|m{\leftcol}@{\hspace{8pt}}|m{\midcola}@{\hspace{8pt}}|m{\midcolb}@{\hspace{8pt}}|m{\midcolc}@{\hspace{8pt}}|m{\rightcol}@{\hspace{8pt}}|}
\hline
\textbf{Network} & \textbf{Description} & \textbf{Layer $A$} & \textbf{Layer $B$} & \textbf{Interlayer edges} 
\\\hline \vspace{10pt}
\begin{tabular}[c]{m{\leftcol}} 
ER--ER same (see Sec.~\ref{sec:ER_same})  \end{tabular} 
& \begin{tabular}{m{\midcola}} Both layers are Erd\H{o}s--R\'enyi (ER) networks from the $G(N,p)$ model with $N = 1000$ nodes and identical connection probability $p$.   \end{tabular} 
& \begin{tabular}{m{\midcolb}} $p_A = p = 10/N$  \end{tabular} 
& \begin{tabular}{m{\midcolc}} $p_B = p = 10/N$ \end{tabular} 
& \begin{tabular}{m{\rightcol}}All nodes connect to their counterparts.  \end{tabular} 
 \\[15pt] \hline
 \begin{tabular}[c]{m{\leftcol}} 
ER--ER heterogeneous (see Sec.~\ref{sec:ER_diff})  \end{tabular} 
& \begin{tabular}{m{\midcola}} Both layers are ER networks from the $G(N,p)$ model with $N=1000$ nodes and different connection probabilities $p_A$ and $p_B$.  \end{tabular} 
& \begin{tabular}{m{\midcolb}} $p_A = 10/N$  \end{tabular}
& \begin{tabular}{m{\midcolc}} $p_B = 15/N$  \end{tabular}
& \begin{tabular}{m{\rightcol}}All nodes connect to their counterparts.  \end{tabular} 
 \\\hline \vspace{20pt}
 \begin{tabular}[c]{m{\leftcol}} 
CP--ER interlayer all (see Sec.~\ref{sec:CP_ER}) \end{tabular} 
& \begin{tabular}{m{\midcola}} Layer $A$ is a 500-node core--periphery (CP) network that we generate from a stochastic block model (SBM), with a core block (25\% of the nodes) and periphery block (75\% of the nodes), and $P_{\textrm{cc}} > P_{\textrm{cp}} > P_{\textrm{pp}}$. Layer $B$ is an ER network with $N = 500$ nodes. \end{tabular}
& \begin{tabular}{m{\midcolb}}
\centering
$\begin{aligned}
P_{\textrm{cc}} &= 30/n_{\textrm{c}} \\
P_{\textrm{cp}} &= 10/N \\
P_{\textrm{pp}} &= 3/n_{\textrm{p}}
\end{aligned}$
\end{tabular}
& \begin{tabular}{m{\midcolc}} $p = 10/N$ \end{tabular} 
& \begin{tabular}{m{\rightcol}} All nodes connect to their counterparts. \end{tabular}
\\[30pt] \hline \vspace{10pt}
\begin{tabular}[c]{m{\leftcol}} 
CP--ER interlayer core (see Sec.~\ref{sec:CP_ER}) \end{tabular}
& \begin{tabular}{m{\midcola}} The same as above, but only core nodes in layer $A$ connect to their counterparts in layer $B$. \end{tabular}
& \begin{tabular}{m{\midcolb}}
\centering
$\begin{aligned}
P_{\textrm{cc}} &= 30/n_{\textrm{c}} \\
P_{\textrm{cp}} &= 10/N \\
P_{\textrm{pp}} &= 3/n_{\textrm{p}}
\end{aligned}$
\end{tabular}
& \begin{tabular}{m{\midcolc}} $p = 10/N$ \end{tabular}
& \begin{tabular}{m{\rightcol}} Core nodes in layer $A$ connect to their counterparts. \end{tabular}
\\[20pt] \hline \vspace{5pt}
\begin{tabular}[c]{m{\leftcol}} CP--ER interlayer peripheral (see Sec.~\ref{sec:CP_ER}) \end{tabular} 
& \begin{tabular}{m{\midcola}} The same as above, but only peripheral nodes in layer $A$ connect to their counterparts in layer $B$. \end{tabular}
& \begin{tabular}{m{\midcolb}}
\centering
$\begin{aligned}
P_{\textrm{cc}} &= 30/n_{\textrm{c}} \\
P_{\textrm{cp}} &= 10/N \\
P_{\textrm{pp}} &= 3/n_{\textrm{p}}
\end{aligned}$
\end{tabular}
& \begin{tabular}{m{\midcolc}} $p = 10/N$ \end{tabular}
& \begin{tabular}{m{\rightcol}} Peripheral nodes in layer $A$ connect to their counterparts. \end{tabular}
\\[20pt]\hline
\begin{tabular}[c]{m{\leftcol}} 
\emph{C. elegans} multiplex connectome (see Sec.~\ref{sec:c_elegans}) \end{tabular}
& \begin{tabular}{m{\midcola}}Multiplex network of synaptic and extrasynaptic signaling between neurons \cite{varshney2011structural,ripoll2023neuropeptidergic}. \end{tabular} 
& \begin{tabular}{m{\midcolb}}  \centering --  \end{tabular}
& \begin{tabular}{m{\midcolc}}  \centering --  \end{tabular}
& \begin{tabular}{m{\rightcol}} All nodes connect to their counterparts. \end{tabular} 
\\\hline
\begin{tabular}[c]{m{\leftcol}} 
UMich--MSU Facebook network (see Sec.~\ref{sec:fb_nets}) \end{tabular}
& \begin{tabular}[c]{m{\midcola}} Facebook friendship network between individuals at University of Michigan and Michigan State University \cite{melnik2014dynamics,traud2012}. \end{tabular}
& \begin{tabular}{m{\midcolb}} \centering -- \end{tabular}
& \begin{tabular}[c]{m{\midcolc}} \centering --  \end{tabular}
& \begin{tabular}{m{\rightcol}} Interlayer edges connect users from different universities. \end{tabular}
\\\hline

\end{tabular}
\end{table*}


\subsection{Numerical simulations}

Researchers typically use MC methods to sample spin configurations and thereby study Ising dynamics numerically~\cite{sethna2021statistical,newman1999monte}. To do this, we employ the Metropolis--Hastings (MH) algorithm \cite{newman1999monte,sandvik-mc} for each value of the temperature $T$. The MH algorithm begins by generating a random initial spin configuration, which we determine by assigning a spin of either $+1$ or $-1$ to each state node with probability $1/2$ for each value. At each time step, one then chooses a state node uniformly at random. If flipping the spin of the chosen state node decreases the energy of the system, we accept the spin flip. Otherwise, we accept the spin flip with probability $w = \exp{(-\Delta E/T)}$, where $\Delta E$ is the difference in energy due to the spin flip. A single iteration of this MC method has $S$ time steps, where $S$ is the total number of spins in a network. To minimize the effects of the initial spin configuration, we discard the first $5 \times 10^{3}$ MC iterations. We then calculate quantities of interest by averaging over subsequent MC iterations. We report means of $5 \times 10^{5}$ iterations after discarding the initial $5 \times 10^{3}$  iterations. 


\medskip
\medskip

\subsection{Mean-field approximation} \label{sec:mean_field_approx}

MC simulations can be computationally expensive for large networks, and they can also be challenging to interpret. It is often instructive to employ approximation schemes, which can provide complementary insights to numerical simulations and allow one to study associated free-energy landscapes. A standard approximation method is a mean-field approximation \cite{sethna2021statistical,kardar2007statistical,goldenfeld2018lectures}, which we use to obtain a self-consistency equation
\begin{equation}
    m_i^{\textrm{MF}} = \tanh \left( \frac{\sum_{j} J_{ij} m_j^{\textrm{MF}}}{T}\right) 
    \label{eq:self_consistency1}
\end{equation}
for the expected value $m_i^{\textrm{MF}}$ of each state node $i$.
We solve {Eq.} \eqref{eq:self_consistency1} using a fixed-point iteration scheme. 
\begin{widetext}
Under this mean-field approximation, the Gibbs free energy is
\begin{equation}
    G_{\textrm{MF}} = - \sum_{i,j} J_{ij} m_i m_j + T \left\{ \sum_{i} \left[ \frac{1 + m_i}{2} \ln \left( \frac{1+m_i}{2} \right) + \frac{1 - m_i}{2} \ln \left( \frac{1 - m_i}{2} \right) \right] \right\} \,.
    \label{eq:free_energy1}
\end{equation}
In Appendix \ref{sec:meanfield_derivation}, we give details of our derivations of Eqns. \eqref{eq:self_consistency1} and \eqref{eq:free_energy1}. 
\end{widetext}

Solutions of Eq.~\eqref{eq:self_consistency1} correspond to stationary points of the Gibbs free energy. These stationary points can be local minima, local maxima, or saddle points. To evaluate the nature of a stationary point, we compute the Hessian 
\begin{equation} \label{eq:hessian1}
     H(G_{\textrm{MF}}) = - J + T \, \textrm{Diag} \left[ \frac{1}{1 - m_i^2}\right]   
\end{equation}
of the Gibbs free energy. A solution is a local minimum of the Gibbs free energy if all eigenvalues of the Hessian \eqref{eq:hessian1} are positive. Equilibrium solutions of the mean-field Ising model \eqref{eq:self_consistency1} correspond to global minima of the Gibbs free energy. However, depending on the network structure and Ising model parameters, the Gibbs free energy can have other local minima, which are metastable states. If we initialize the spins at such a local minimum or in the basin of attraction of a local minimum, then the system can remain trapped in that basin for a long time~\cite{krapivsky2010kinetic}. 

In the Ising model on coupled networks or on networks with community structure (i.e., modular networks), metastable states can manifest as distinct clusters with magnetizations that have opposite signs \cite{bolfe2018ising_coupled,dasgupta2009phase,suchecki2009bistable}. Similar phenomena have been observed in models of opinion dynamics on networks with community structure, where different communities of nodes have different opinions \cite{peng2022majority,lambiotte2007coexistence,castellano2009statistical}. Metastable states also arise in the context of ``freezing" in Ising ferromagnets \cite{spirin2001freezing,krapivsky2010kinetic} and more generally in spin glasses \cite{zamponi2010mean,castellani2005spin}.

Although our mean-field approximation does not give quantitatively accurate results for the examined networks, especially near critical points, it can capture key qualitative features of phase diagrams and it enables analysis of free-energy landscapes. It thereby allows us to develop intuition for Ising dynamics on multiplex networks. Additionally, the self-consistency equation \eqref{eq:self_consistency1} is itself of interest because one can view the set of these equations as a neuronal system in which each spin $i$ represents the firing rate of a neuron with a logistic activation function~\cite{lynn2024heavy}. It is important to consider the assumptions that we make in our mean-field approximation, as these assumptions can inform its limitations. In particular, our approximation factorizes the joint probability distribution of spin configurations into a product of independent single-spin distributions. See Eq.~\eqref{eq:trial_prob} in Appendix~\ref{sec:meanfield_derivation}. This simplification neglects correlations between the spins of state nodes, and it tends to perform poorly when there are significant fluctuations around their mean values. For a $d$-dimensional hypercubic lattice, the above mean-field approximation becomes increasingly reliable as one increases the dimension (and is generally considered to be accurate for $d \ge 4$), but it performs poorly in low dimensions, erroneously predicting a phase transition for $d = 1$ \cite{goldenfeld2018lectures}. As we observe in Section {\ref{sec:ER_mono}}, this mean-field approximation tends to work better for networks with large mean degrees. Previous studies on other dynamical processes on networks have demonstrated that mean-field approximations can also be accurate when the mean first-neighbor degree (i.e., the mean degree of the neighbors of a network's nodes) of a network is sufficiently large \cite{accuracy2012}. For each of the synthetic networks that we consider, we evaluate the accuracy of our mean-field approximation by comparing it to simulations for many values of the network parameters.


\section{Case study with synthetic networks: Erd\H{o}s--R\'enyi (ER) networks} \label{sec:ER_graphs}


To build intuition for the Ising Hamiltonian \eqref{eq:hamiltonian}, we first examine it on networks in which each layer is an ER network. We study the steady-state behavior of the overall and layer magnetizations as functions of the temperature $T$ and interlayer coupling $r$ using both MC simulations and the mean-field approximation in Section \ref{sec:mean_field_approx}. We track the mean absolute magnetization of both the overall multiplex network and each individual layer. Henceforth, we use the term ``mean magnetization" in our MC simulations to refer specifically to the mean absolute magnetization. We generate ER networks using the $G(N,p)$ model, where $N$ is the number of nodes in a layer (i.e., the layer size) and we include each possible edge between two state nodes with independent probability $p$. To ensure that all ER networks are connected, we consider $p > \ln N/N$ \cite{newman2018networks}. We also ensure numerically that all of the networks that we generate are connected.

Before examining multiplex ER networks, we compare our mean-field approximation to MC simulations on monolayer ER networks. We then construct a two-layer multiplex network with ER layers with the same values of $p$ and $N$. Finally, to study the effects of heterogeneous layer densities, we consider heterogeneous connection probabilities $p_A$ and $p_B$ in the two layers.


\subsection{Ising dynamics on monolayer ER networks} \label{sec:ER_mono}

\begin{figure}
    \centering
    \includegraphics[width=0.85\linewidth]{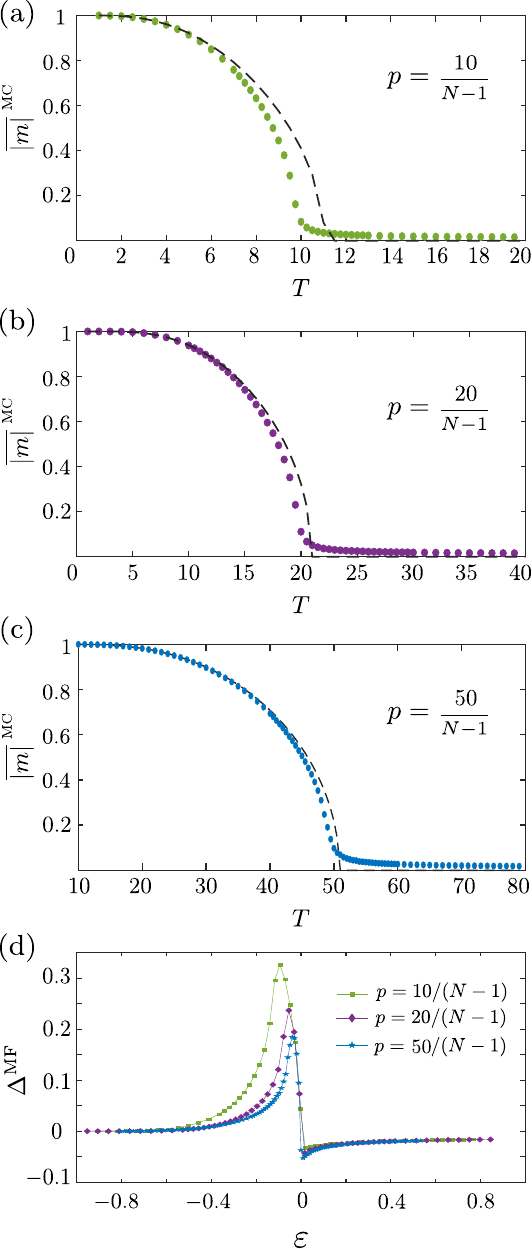}
    \caption{Comparison between Monte Carlo (MC) simulations and our mean-field approximation of Ising dynamics on ER networks with $N = 5000$ nodes. We show the mean absolute magnetization as a function of temperature for ER networks with connection probabilities (a) $p = 10/(N - 1)$, (b) $p = 20/(N - 1)$, and (c) $p = 50/(N - 1)$. In panel (d), we show the approximation error $\Delta^{\textrm{MF}}$ as a function of the reduced temperature $\varepsilon$. In each panel, the markers indicate the means of MC simulations and dashed curves indicate mean-field values. For each temperature, we report means of MC results over $5 \times 10^5$ MC iterations after discarding the first $5 \times 10^3$ MC iterations.}  \label{fig:ER_singlelayer}
\end{figure}

We construct monolayer ER networks with $N = 5000$ nodes and connection probabilities $p = 10/(N - 1)$, $p = 20/(N - 1)$, and $p = 50/(N - 1)$. The resulting networks have expected mean degrees of $\langle k \rangle \approx 10$, $\langle k \rangle \approx 20$, and $\langle k \rangle \approx 50$, respectively. We examine Ising dynamics on these networks and assess the accuracy of our mean-field approximation as a function of mean degree. In Fig.~\ref{fig:ER_singlelayer}, we summarize the results of MC simulations and our mean-field approximation. As expected, we observe that the temperature at which the magnetization of each network approaches $0$ increases with the mean degree. At high temperatures, the magnetization in numerical simulations does not drop precisely to $0$ due to finite-size effects \cite{privman1990finite}. Moreover, although our mean-field approximation captures qualitative trends, it deviates from MC simulations near the critical temperature, highlighting its limitations in capturing the critical behavior.

To quantify deviations between our mean-field approximation and MC simulations for different values of the connection probability $p$, we calculate
\begin{equation}
    \Delta^{\textrm{MF}} = m^{\textrm{MF}} - \overline{m}^{\textrm{MC}},
\end{equation}
where $m^{\textrm{MF}}$ is the magnetization from the  mean-field approximation and $\overline{m}^{\text{MC}}$ is the magnetization from MC simulations.

To compare the mean-field and MC magnetizations on equal footing, we calculate a ``reduced temperature"
\begin{equation}
    \varepsilon = \frac{T_c^{\textrm{MF}} - T}{T_c^{\textrm{MF}}}\,,
\end{equation}
where $T_c^{\textrm{MF}}$ is the mean-field critical temperature. The reduced temperature is a normalized measure of proximity to the mean-field critical temperature. In Fig.~\ref{fig:ER_singlelayer}(d), we calculate the deviation $\Delta^{\textrm{MF}}$ as a function of $\varepsilon$. For a fixed network size $N$, we observe that this approximation error decreases as we increase the connection probability $p$. This observation aligns with the general intuition that a mean-field approximation becomes more accurate as one increases the mean number of neighbors of the nodes (i.e., the mean degree) of a network~\cite{accuracy2012}.


\subsection{Two-layer multiplex networks with ER layers with the same connection probability}\label{sec:ER_same}

\begin{figure*}
    \centering
    \includegraphics[width=0.9\linewidth]{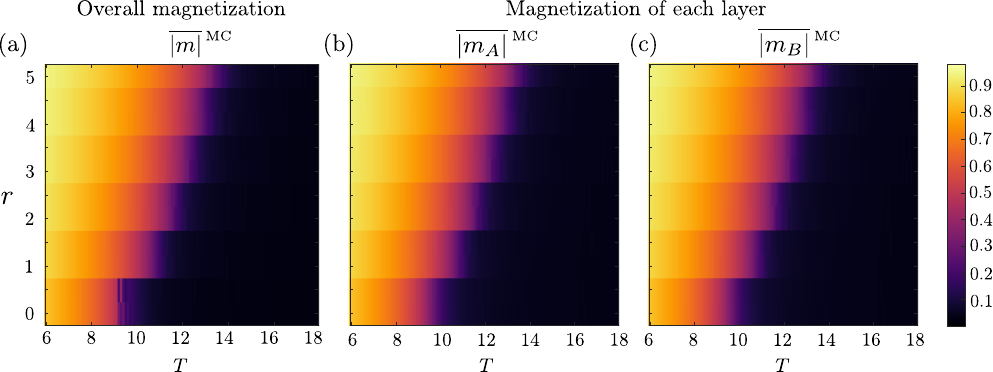}
    \caption{Ising dynamics on a two-layer multiplex network in which the layers are ER networks with the same connection probability. We plot heat maps of (a) the mean absolute magnetization $\overline{|m|}^{\textrm{MC}}$ of the overall multiplex network, (b) the mean absolute magnetization $\overline{|m_{A}|}^{\textrm{MC}}$ of layer $A$, and (c) the mean absolute magnetization $\overline{ |m_{B}|}^{\textrm{MC}}$ of layer $B$. In each heat map, the vertical axis indicates the interlayer coupling strength $r$ and the horizontal axis indicates the temperature $T$. For each temperature, we report means of MC results over $5 \times 10^5$ MC iterations after discarding the first $5 \times 10^3$ MC iterations. Each layer has $N = 1000$ nodes (resulting in a multiplex network with $S = 2000$ state nodes) and a connection probability of $p = 10/(N - 1)$. }
    \label{fig:MC_ER_k10k10}
\end{figure*}

We now consider a multiplex network in which each layer is an ER $G(N,p)$ network with identical connection probability $p$ and size $N$. The layers have $N = 1000$ state nodes and a connection probability of $p = 10/(N - 1)$, resulting in an expected mean degree of $\langle k \rangle \approx 10$ for each layer. We combine the two layers into a multiplex network, and we use MC simulations to examine Ising dynamics as a function of the temperature $T$ for different values of the interlayer coupling strength $r$. In Fig.~\ref{fig:MC_ER_k10k10}, we show results using a single instantiation of each ER layer. We observe the same qualitative results for other instantiations. We observe that the heat maps of the mean magnetization of the overall multiplex network and of each layer are identical.  At low temperatures, both the layer magnetization and overall magnetization are nonzero; at high temperatures, the magnetizations approach $0$. As we increase the interlayer coupling strength $r$, there is an increase in the temperature at which there is a transition between ordered and disordered states, as strengthening the interlayer coupling enhances the correlations between state nodes across layers. 

\begin{figure*}
    \centering
    \includegraphics[width=\linewidth]{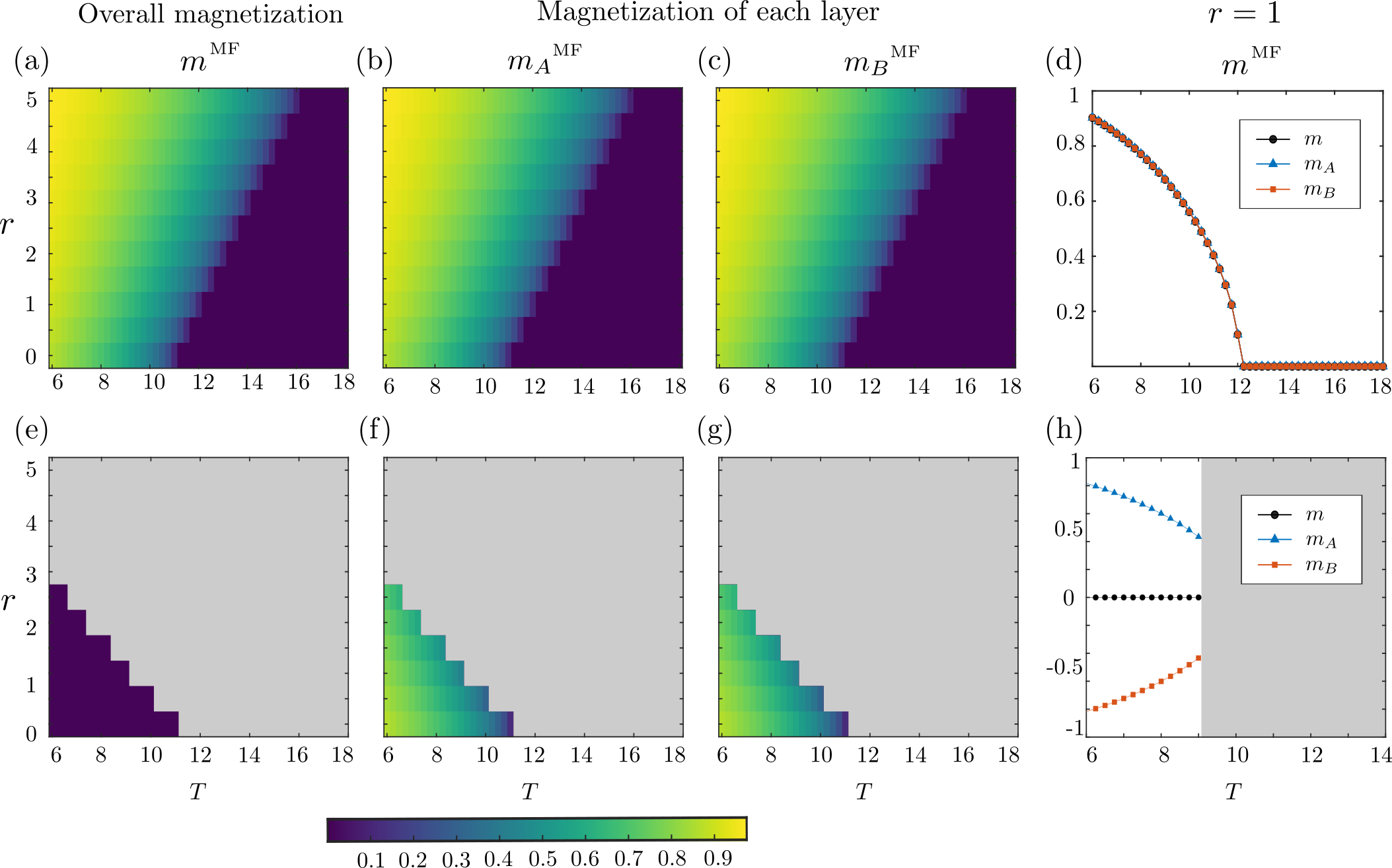}
    \caption{Mean-field Ising dynamics on a two-layer multiplex network in which the layers are ER networks with the same connection probability. In (a)--(d), we show a solution in which both layers have magnetizations with the same sign. In (e)--(h), we show a metastable anti-aligned solution, in which the two layers have magnetizations of opposite signs. We plot (a, e) a heat map of the mean-field magnetization $m^{\textrm{MF}}$ of the overall multiplex network, (b, f) a heat map of the mean-field magnetization $m^{\textrm{MF}}_{A}$ of layer $A$, (c, g) a heat map of the mean-field magnetization $m^{\textrm{MF}}_{B}$ of layer $B$, and (d, h) the mean-field magnetizations of the overall network and of both layers as functions of temperature $T$ for an interlayer coupling strength of $r = 1$. In each heat map, the vertical axis indicates the interlayer coupling strength $r$ and the horizontal axis indicates the temperature $T$. In panels (e)--(h), the gray regions indicate locations in which {we do not observe a metastable anti-aligned solution for the examined initial conditions.} Each layer has $N = 1000$ state nodes and a connection probability of $p = 10/(N - 1)$, resulting in a two-layer multiplex network with $S = 2000$ state nodes. 
}
\label{fig:MF_ER_k10k10}
\end{figure*}

We also examine solutions of the mean-field approximation \eqref{eq:self_consistency1} that we described in Sec. \ref{sec:mean_field_approx}. We first obtain solutions of Eq.~\eqref{eq:self_consistency1} using a fixed-point iteration starting with an initial magnetization of $m_i = 0.5$ for each state node. We show our results in the top row of Fig.~\ref{fig:MF_ER_k10k10}. In Figs.~\ref{fig:MF_ER_k10k10}(a)--(c). we show heat maps of the mean-field magnetizations of the overall network, layer $A$, and layer $B$. In Fig.~\ref{fig:MF_ER_k10k10}(d), we show the mean-field overall and layer magnetizations for interlayer coupling strength $r = 1$. These results are qualitatively similar to our MC simulations, but they differ quantitatively. 

As we discussed in Sec.~\ref{sec:mean_field_approx}, the Gibbs free energy may also have local minima that correspond to metastable states. We explore the possibility of metastable states in which the two layers have nonzero magnetizations of opposite signs. We refer to such a configuration as ``anti-aligned". An anti-aligned configuration is not an equilibrium solution, but we expect it to be metastable for small values of temperature $T$ and interlayer coupling strength $r$. If we initialize the system in or sufficiently near an anti-aligned state, it can remain or become trapped in this local minimum. To study the presence of an anti-aligned configuration, we compute fixed-point solutions of Eq.~\eqref{eq:self_consistency1} for an initial condition with $m_i = 1$ for each state node in layer $A$ and $m_i = -1$ for each state node in layer $B$. We then study whether or not such a solution is a local minimum by checking if the Hessian is positive definite. In Figs.~\ref{fig:MF_ER_k10k10}(e)--(h), we show the resulting mean-field overall and layer magnetizations. For small interlayer coupling strengths $r$, we observe that the anti-aligned state is metastable until we increase $T$ to a particular temperature. Beyond this temperature, we no longer obtain an anti-aligned solution that is a minimum of the free energy. As we increase $r$, we observe a decrease in the temperature range for which an anti-aligned solution is metastable. 

\subsection{Two-layer multiplex networks with ER layers with the different connection probabilities}\label{sec:ER_diff}

\begin{figure*}
    \centering
    \includegraphics[width=0.9\linewidth]{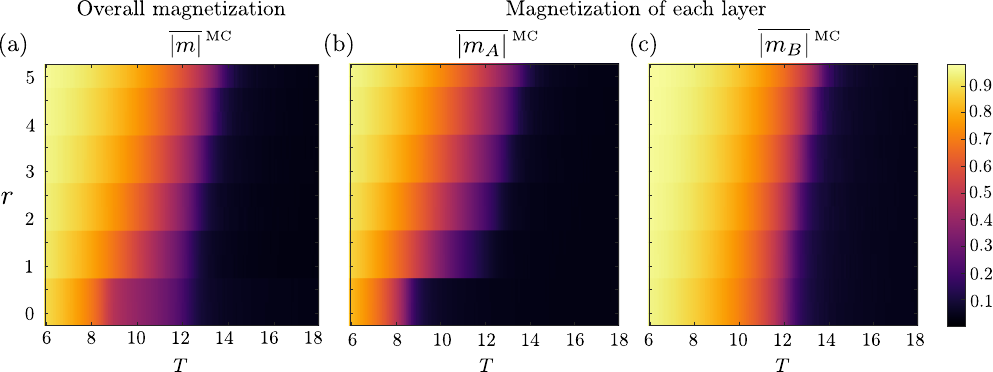}
    \caption{Ising dynamics on a two-layer multiplex network in which the layers are ER networks with different connection probabilities. We plot heat maps of (a) the mean absolute magnetization $\overline{|m|}^{\textrm{MC}}$ of the overall multiplex network, (b) the mean absolute magnetization $\overline{|m_{A}|}^{\textrm{MC}}$of layer $A$, and (c) the mean absolute magnetization $\overline{|m_{B}|}^{\textrm{MC}}$ of layer $B$. In each heat map, the vertical axis indicates the interlayer coupling strength $r$ and the horizontal axis indicates the temperature $T$. For each temperature, we the MC results are means of $5 \times 10^5$ MC iterations after discarding the first $5 \times 10^3$ MC iterations. Each layer has $N = 1000$ state nodes (resulting in a multiplex network with $S = 2000$ state nodes), a connection probability of $p = 10/(N - 1)$ in layer $A$, and a connection probability of $p = 15/(N - 1)$ in layer $B$.
 }
    \label{fig:MC_ER_k10k15}
\end{figure*}

Building on our previous calculations, we now examine Ising dynamics on a two-layer multiplex network in which one layer is denser than the other. We consider a network that consists of ER networks with $N = 1000$ state nodes in each layer, a connection probability of $p = 10/(N - 1)$ on layer $A$, and a connection probability of $p = 15/(N - 1)$ on layer $B$. These layers have expected mean degrees of approximately 10 and 15, respectively. We report results on simulations of a single instantiation of this two-layer multiplex network, but we observe the same qualitative results for different instantiations. 

In Fig.~\ref{fig:MC_ER_k10k15}, we show the results of MC simulations of the Ising {dynamics}. We observe that the overall magnetization of the multiplex network approaches $0$ at a higher temperature than for ER layers with equal densities (see Fig.~\ref{fig:MC_ER_k10k10}). We also observe that the overall magnetization of the network now achieves an intermediate value (specifically, $\langle |m| \rangle \in [0.3, 0.6])$) for a wider range of temperatures in Fig.~\ref{fig:MC_ER_k10k15} than in Fig.~\ref{fig:MC_ER_k10k10}. By examining the individual magnetizations of  each layer, especially when the interlayer coupling strength is $r = 0$, we see that this observation arises because the layer with the smaller mean degree (i.e., the sparser layer) loses order at lower values of temperature than the layer with the larger mean degree (i.e., the denser layer). Upon increasing $r$, the sparser layer retains a small amount of magnetization past its own critical temperature, as it is now coupled to the denser network (which still has long-range order).

\begin{figure*}
    \centering
    \includegraphics[width=\linewidth]{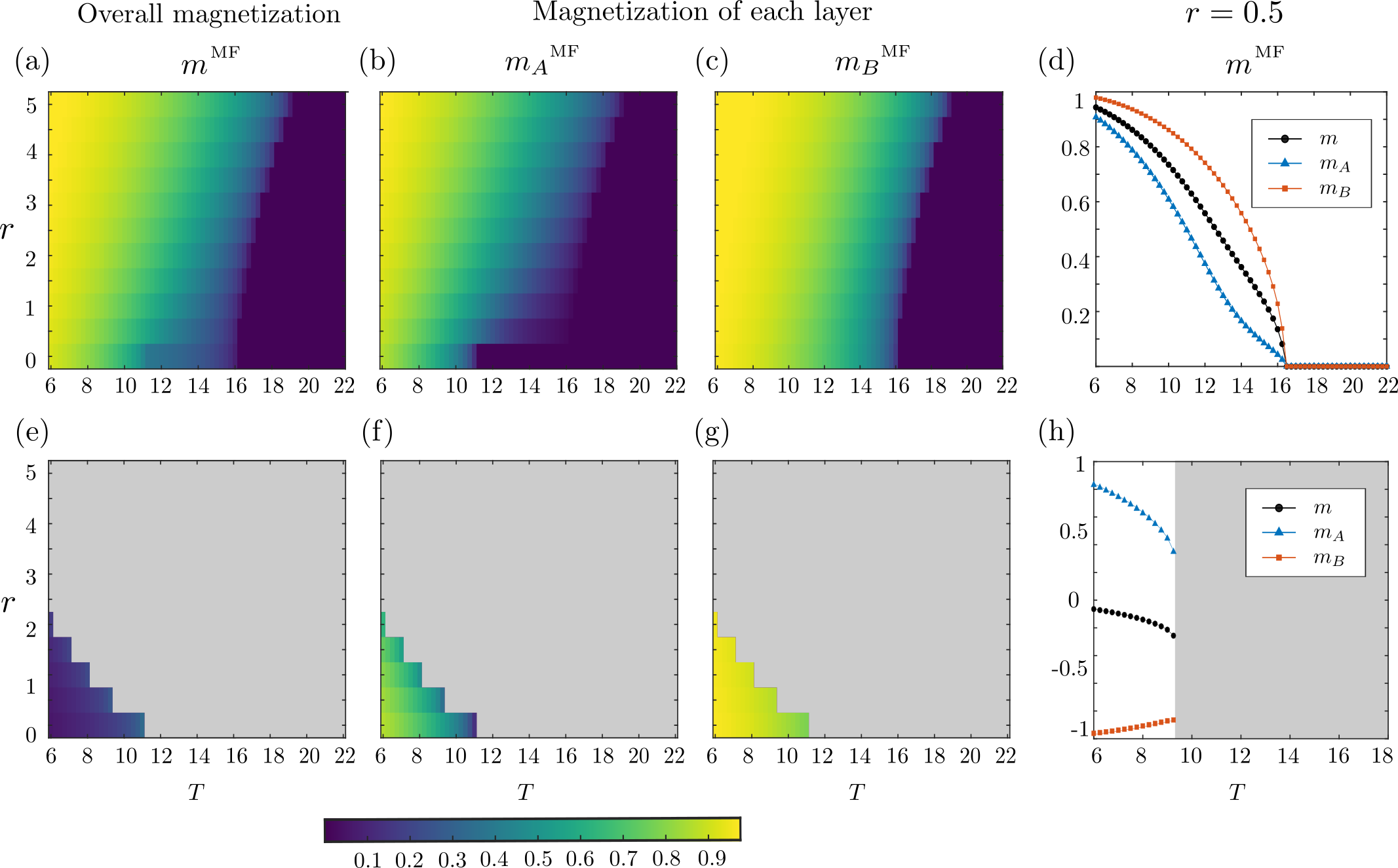}
    \caption{Mean-field Ising dynamics on a two-layer multiplex network in which the layers are ER networks with different connection probabilities. In (a)--(d), we show a solution in which both layers have magnetizations with the same sign. In (e)--(h), we show a metastable anti-aligned solution, in which the two layers have magnetizations of opposite signs. We plot (a, e) a heat map of the mean-field magnetization $m^{\textrm{MF}}$ of the overall multiplex network, (b, f) a heat map of the mean-field magnetization $m^{\textrm{MF}}_{A}$ of layer $A$, (c, g) a heat map of the mean-field magnetization $m^{\textrm{MF}}_{B}$ of layer $B$, and (d, h) the mean-field magnetizations of the overall network and of both layers as functions of temperature $T$ for an interlayer coupling strength of $r = 0.5$. In each heat map, the vertical axis indicates the interlayer coupling strength $r$ and the horizontal axis indicates the temperature $T$. {In panels (e)--(h), the gray regions indicate locations in which we do not observe a metastable anti-aligned solution for the examined initial conditions.} Each layer has $N = 1000$ nodes, resulting in a multiplex network with $S = 2000$ state nodes. The connection probability is $p = 10/(N - 1)$ in layer $A$ and $p = 15/(N - 1)$ in layer $B$. }
    \label{fig:MF_ER_k10k15}
\end{figure*}

We also study our mean-field approximation of the Ising dynamics on this two-layer network. We compute fixed-point solutions of Eq.~\eqref{eq:self_consistency1} starting with an initial condition of $m_i = 0.5$ for each state node $i$. The resulting mean-field magnetizations, which we show in Figs.~\ref{fig:MF_ER_k10k15}(a)--(d), are qualitatively consistent with the results of our MC simulations. To examine the presence of metastable anti-aligned steady states, we obtain fixed-point solutions of Eq.~\eqref{eq:self_consistency1} {for} an initial condition with $m_i = 1$ for state nodes in layer $A$ and $m_i = -1$ for state nodes in layer $B$. We show the results of this computation in Figs.~\ref{fig:MF_ER_k10k15}(e)--(h). Because one layer is denser than the other, we obtain different magnetizations in the two layers and also obtain a nonzero overall magnetization. Additionally, as we increase the interlayer coupling strength $r$, we observe that the temperature range over which the anti-aligned configuration is metastable shrinks more than in the equal-density case (see Fig.~\ref{fig:MF_ER_k10k10}). 

%


\section{Case study with synthetic networks: Core--periphery (CP) structure on one layer}\label{sec:CP_graphs}

Many real networks have community structure and/or other mesoscale features~\cite{newman2018networks}. One important mesoscale feature is CP structure \cite{rombach2017core,csermely2013structure}. CP structure is relevant in social networks \cite{boorman1976social}, transportation networks \cite{lee2014density}, brain networks \cite{bassett2013task,kulkarni2025toward}, and many other types of networks. We consider the simplest type of CP structure, which consists of a single set of core nodes and a single set of peripheral nodes. The core nodes are densely interconnected and are reasonably well-connected to peripheral nodes, and the peripheral nodes are sparsely connected both to core nodes and to each other. Let $n_{\textrm{core}}$ denote the number of core nodes (i.e., the size of the core set), and let $n_{\textrm{peri}}$ denote the number of peripheral nodes (i.e., the size of the periphery set). We generate synthetic CP networks using a stochastic block model (SBM) \cite{rombach2017core, csermely2013structure}. The edge-probability matrix of this SBM matrix is
\begin{equation}
    P = \begin{pmatrix}
         P_{\textrm{cc}} & P_{\textrm{cp}} \\
         P_{\textrm{cp}} & P_{\textrm{pp}} 
        \end{pmatrix}\,,
\end{equation}
where $P_{\textrm{cc}}$, $P_{\textrm{cp}} = P_{\textrm{pc}}$, and $P_{\textrm{pp}}$ denote the probabilities of constructing edges between core nodes, core and peripheral nodes, and peripheral nodes, respectively. These probabilities satisfy $P_{\textrm{cc}} \geq P_{\textrm{cp}} \geq P_{\textrm{pp}}$. Because we consider undirected networks, $P_{\textrm{cp}} = P_{\textrm{pc}}$. Choosing the probabilities $P_{\textrm{cc}} =  P_{\textrm{cp}} = P_{\textrm{pp}}$ reduces a CP network to an ER network, which does not have actual CP structure.

\subsection{Ising dynamics on monolayer networks with CP structure}

Before incorporating CP structure into multiplex networks, we consider Ising dynamics on monolayer networks with CP structure. Chen et al.~\cite{chen2018double} studied the ferromagnetic Ising model on CP networks with a single core set and a single periphery set. They observed a conventional order--disorder transition, with networks in a disordered state above a critical temperature. However, for a range of temperatures below this order--disorder transition, they identified a state in which the core set is more ordered than the periphery set. The magnetic susceptibility $\chi = N\! \left[\langle m^2 \rangle - \langle m \rangle^2 \right]\!/T$ has two distinct peaks as a function of temperature $T$. Depending on how the number of edges between core nodes and peripheral nodes scales with the system size $N$, the heights of both peaks increase with $N$ and become infinite as $N \rightarrow \infty$~\cite{chen2018double}. Double phase transitions have also been examined in other dynamical processes on modular networks \cite{colomer2014double} and in asymmetric percolation processes \cite{allard2017asymmetric}.

\begin{figure}
    \centering\includegraphics[width=0.95\linewidth]{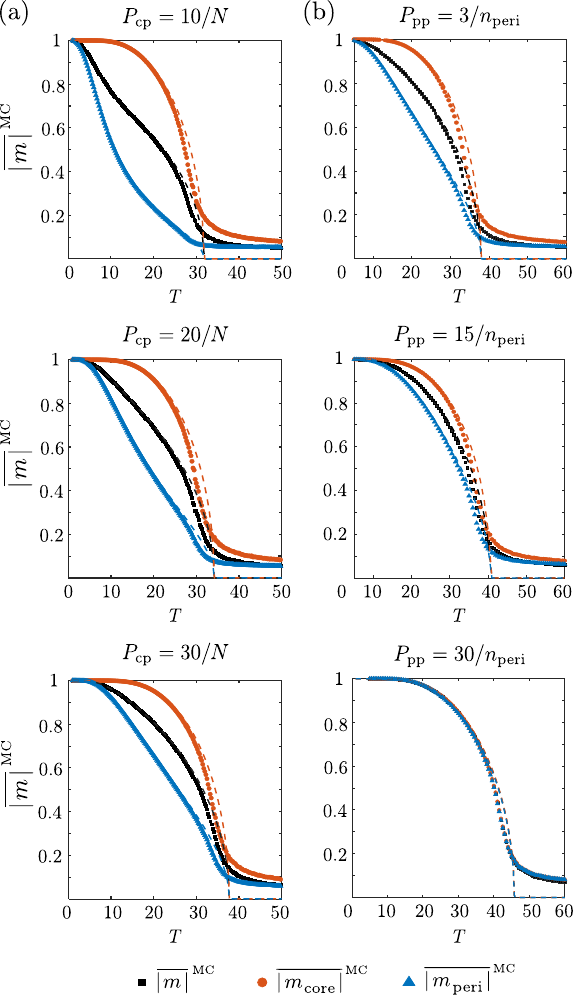}
    \caption{Comparison between our MC simulations and the mean-field approximation of the Ising dynamics
    on monolayer networks with CP structure. We show the mean magnetizations of the overall network, the core nodes, and the peripheral nodes as functions of temperature. We explore (a) the effects of varying the connection probability $P_{\textrm{cp}}$ between core nodes and peripheral nodes and (b) the effects of varying the connection probability $P_{\textrm{pp}}$ between peripheral nodes. The markers indicate means of MC simulations, and the dashed curves indicate mean-field values. For each temperature, the MC results are means of $5 \times 10^5$ MC iterations after discarding the first $5 \times 10^3$ MC iterations. Each network has $N = 500$ nodes, with $n_{\textrm{core}} = 250$ core nodes and $n_{\textrm{peri}} = 250$ peripheral nodes.}
    \label{fig:CP_single}
\end{figure}

We examine magnetization as a function of the SBM parameters. We consider networks with $N = 500$ nodes in which the core set and periphery set have the same size, so $n_{\textrm{core}} = n_{\textrm{peri}} = 250$. Motivated by an example in Ref.~\cite{kureh2020fitting}, we use the connection probabilities $P_{\textrm{cc}} = 30/n_{\textrm{core}}$, $P_{\textrm{cp}} = 10/N$, and $P_{\textrm{pp}} = 3/n_{\textrm{peri}}$. We monitor the mean magnetizations of the overall network, the core set, and the periphery set using MC simulations and our mean-field approximation. We denote these mean magnetizations by $\overline{m}$, $\overline{m}_{\textrm{core}}$, and $\overline{m}_{\textrm{peri}}$, respectively. Consistent with the results of Chen et al.~\cite{chen2018double}, we observe [see the top row of Fig. \ref{fig:CP_single}(a)] that $\overline{m}_{\textrm{peri}}$ decreases more rapidly than $\overline{m}_{\textrm{core}}$ as we increase the temperature $T$ and that there is a range of temperatures for which the core set is more ordered than the periphery set. Intuitively, this difference in behavior between the core set and periphery set arises because the core nodes are more densely connected than the peripheral nodes, which results in stronger correlations between the states of two core nodes than between the states of two peripheral nodes. As we increase the temperature $T$, thermal fluctuations disrupt spin alignment more in the periphery set than in the core set. Therefore, the core set is more ordered than the periphery set.

Beyond a critical temperature $T_c$, both $\overline{m}_{\textrm{peri}}$ and $\overline{m}_{\textrm{core}}$ (and hence also $\overline{m}$) approach $0$, so the multiplex network is in a disordered (i.e., paramagnetic) state. The overall magnetization of the network has two distinct drops in value. The first drop in the overall magnetization occurs because the magnetization of the peripheral nodes decreases more rapidly than the magnetization of the core nodes, and the second drop in the overall magnetization occurs when the entire network loses order.

We investigate how the Ising dynamics depends on the SBM parameters in two ways. First, we vary the connection probability $P_{\textrm{cp}}$ between the core and peripheral nodes for fixed values of the other parameters. Second, we vary the connection probability $P_{\textrm{pp}}$ between peripheral nodes for fixed values of the other parameters. To examine the influence of the density of the edges between core and peripheral nodes, we fix $P_{\textrm{cc}} = 30/n_{\textrm{core}}$ and $P_{\textrm{pp}} = 3/n_{\textrm{peri}}$ and consider $P_{\textrm{cp}} = 10/N$, $P_{\textrm{cp}} = 20/N$, and $P_{\textrm{cp}} = 30/N$. In Fig. \ref{fig:CP_single}(a), we observe that the periphery set is ordered for a larger range of temperatures as we increase the probability $P_{\textrm{cp}}$. Accordingly, we do not observe a sharp drop in $\overline{m}$. To examine the influence of the density of the edges between peripheral nodes, we fix $P_{\textrm{cc}} = 30/n_{\textrm{core}}$ and $P_{\textrm{cp}} = 30/N$ and consider $P_{\textrm{pp}} = 3/n_{\textrm{peri}}$, $P_{\textrm{pp}} = 15/n_{\textrm{peri}}$, and $P_{\textrm{pp}} = 30/n_{\textrm{peri}}$. In Fig.~\ref{fig:CP_single}(b), we observe that increasing the probability $P_{\textrm{pp}}$ reduces the disparity in magnetization between the core set and the periphery set. Additionally, in Fig.~\ref{fig:CP_single}, we observe that our mean-field approximation appears to be more accurate for peripheral nodes then for core nodes, especially for small values of $P_{\textrm{cp}}$ and $P_{\textrm{pp}}$. We also consider $P_{\textrm{pp}} = P_{\textrm{cp}} =  P_{\textrm{cc}}$, which reduces to the ER case, for which there is no distinction between the core and periphery sets.

%


\subsection{Two-layer multiplex networks with one CP layer and one ER layer}\label{sec:CP_ER}

We now incorporate CP structure into one layer of a multiplex network. We consider a two-layer multiplex network with $N = 500$ physical nodes. We impose CP structure in layer $A$, with $1/4$ of its state nodes in the core (i.e., $n_{\textrm{core}} = 125$). The parameters of the SBM for layer $A$ are given by the matrix
\begin{equation}
    P = \begin{pmatrix}
         30/n_{\textrm{core}} & 10/N \\
         10/N & 3/n_{\textrm{peri}} 
        \end{pmatrix}\,.
    \label{eq:CP_params}
\end{equation}
Layer $B$ is an ER network with connection probability $p = 10/N$. We consider a single instantiation of each layer for our analysis, but we observe the same qualitative results for different instantiations. The mean degree of layer $A$ is $k_{A} \approx 13$, and the mean degree of layer $B$ is $k_{B} \approx 10$.
In Layer $A$, the mean degree of core nodes is $k_{A, \textrm{core}} \approx 37$ and the mean degree of peripheral nodes is $k_{B,\textrm{peri}} \approx 5$. Accordingly, layers $A$ and $B$ differ both in their densities and in their mesoscale structures. We study Ising dynamics on this two-layer multiplex network as a function of the interlayer correlation strength $r$ and the temperature $T$.

\begin{figure*}
    \centering
    \includegraphics[width=\linewidth]{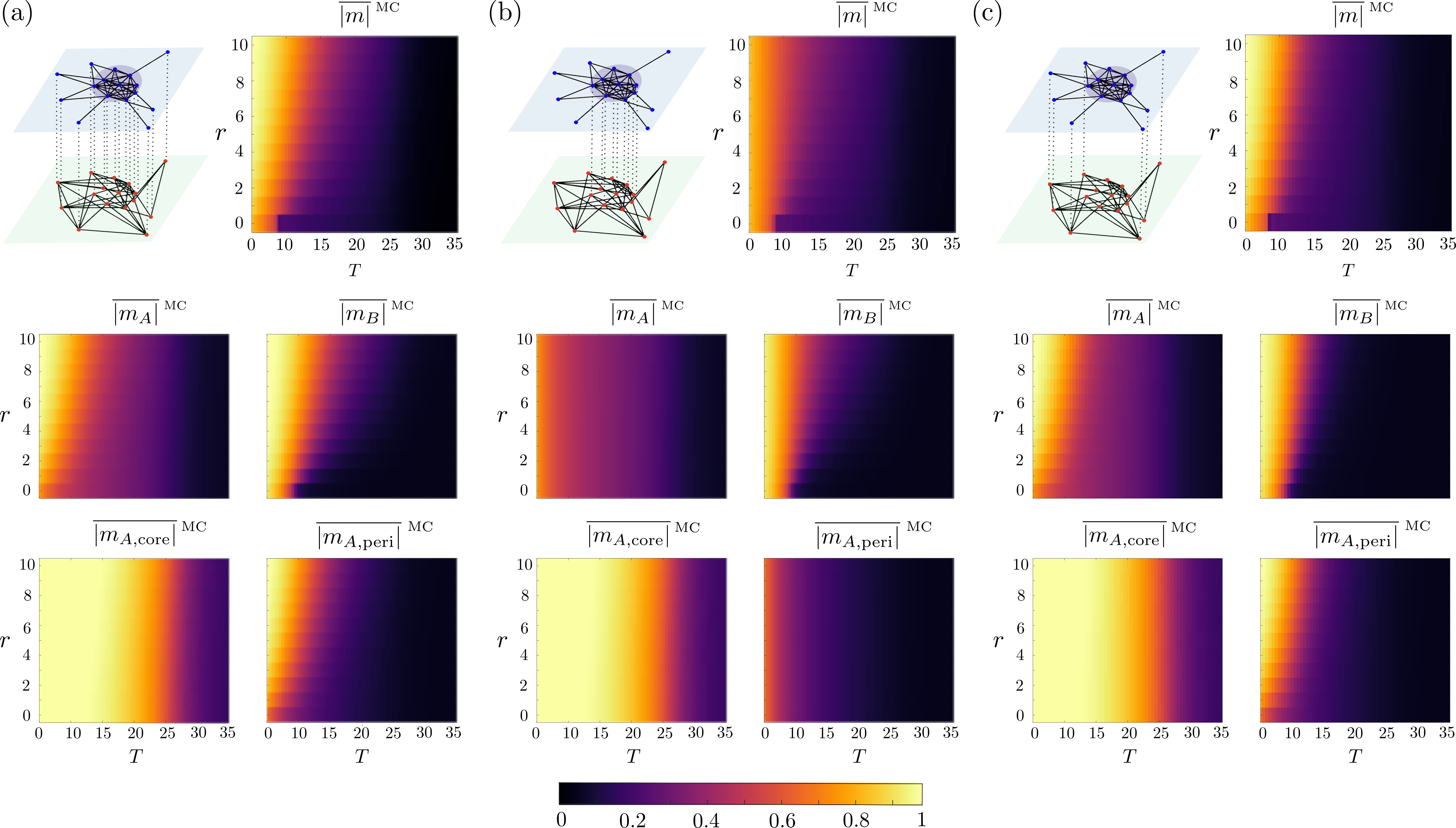}
    \caption{Ising dynamics on a two-layer multilayer network in which layer $A$ has CP structure and layer $B$ is an ER network. We {show} example configurations in which (a) interlayer edges connect all state nodes in layer $A$ with their associated state node in layer $B$, (b) interlayer edges connect only core states nodes in layer $A$ with their corresponding state node in layer $B$, and (c) interlayer edges connect only peripheral state nodes in layer $A$ with their corresponding state node in layer $B$. In all three panels, the top row shows (left) a schematic illustration of the examined multiplex network and (right) the mean overall magnetization of the multiplex network, the middle row shows (left) the mean magnetization of layer $A$ and (right) the mean magnetization of layer $B$, and the bottom row shows (left) the mean magnetization of the core nodes in layer $A$ and (right) the mean magnetization of the peripheral nodes in layer $A$. In each heat map, the horizontal axis indicates the temperature $T$ and the vertical axis indicates the interlayer coupling strength $r$. For each temperature, the MC results are means of $5 \times 10^5$ MC iterations after discarding the first $5 \times 10^3$ iterations.
    }
    \label{fig:CP_ER}
\end{figure*}

We track the overall magnetization, the single-layer magnetizations of layers $A$ and $B$, and the magnetizations of the core set and periphery set in layer $A$. In Fig.~\ref{fig:CP_ER}, we show schematic illustrations of the multiplex networks and the results of MC simulations. As we increase the interlayer coupling strength $r$, we observe that the overall network and each individual layer sustain higher values of magnetization for larger values of temperature. However, in layer $A$, we observe that the core is relatively unaffected by the increase in $r$ for the examined range of $r$. The core nodes have a large mean degree $k_{A,\textrm{core}} \approx 37$, so the dense intralayer connectivity dominates and the comparatively weak interlayer coupling has little effect on these nodes.

The distinct behaviors of the core set and periphery set in layer $A$ motivate us to examine heterogeneities in interlayer edges. We consider two settings. In one setting, we suppose that there are interlayer edges between state nodes if and only if the associated physical node is in the core set in layer $A$ [see Fig.~\ref{fig:CP_ER}(b)]. In the other setting, we suppose that there are interlayer edges between state nodes if and only if the associated physical node is in the periphery set in layer $A$ [see Fig. \ref{fig:CP_ER}(c)]. In both settings, we observe that the core set remains largely unaffected by the interlayer edges. However, the magnetization of the periphery set is larger when peripheral nodes have interlayer edges than when they do not have them. This increases the magnetization of layer $A$ and also enhances the overall magnetization of the multiplex network. It seems that interlayer edges that involve peripheral nodes exert more influence on the system's magnetization than interlayer edges that involve core nodes. A natural extension of our investigation is to examine settings that include nodes that belong to the core set of one layer but to the periphery set of another layer. We expand on this idea and discuss other extensions of our work in Sec.~\ref{sec:discussion}.

%


\section{Ising dynamics on real-world multilayer networks} \label{sec:real_nets}

We now examine multilayer networks that we construct from empirical data. We consider one example of a biological network and one example of a social network. Our biological example is a two-layer multiplex network of distinct signaling modes in the nematode \emph{C. elegans}~\cite{varshney2011structural, ripoll2023neuropeptidergic}. Our social example is a multi-university Facebook network~\cite{melnik2014dynamics,traud2012}, which has a more general multilayer structure than a multiplex network. In both examples, the individual layers have distinct, heterogeneous network structures. In these examples, we compare the Ising dynamics on individual layers to the overall Ising dynamics on the multilayer networks. We thereby gain insight into how empirical multilayer network structures influence dynamical processes.


\subsection{Multiplex connectome of \emph{Caenorhabditis elegans}: Synaptic and extrasynaptic signaling}\label{sec:c_elegans}

\begin{figure*}
    \centering
    \includegraphics[width=\linewidth]{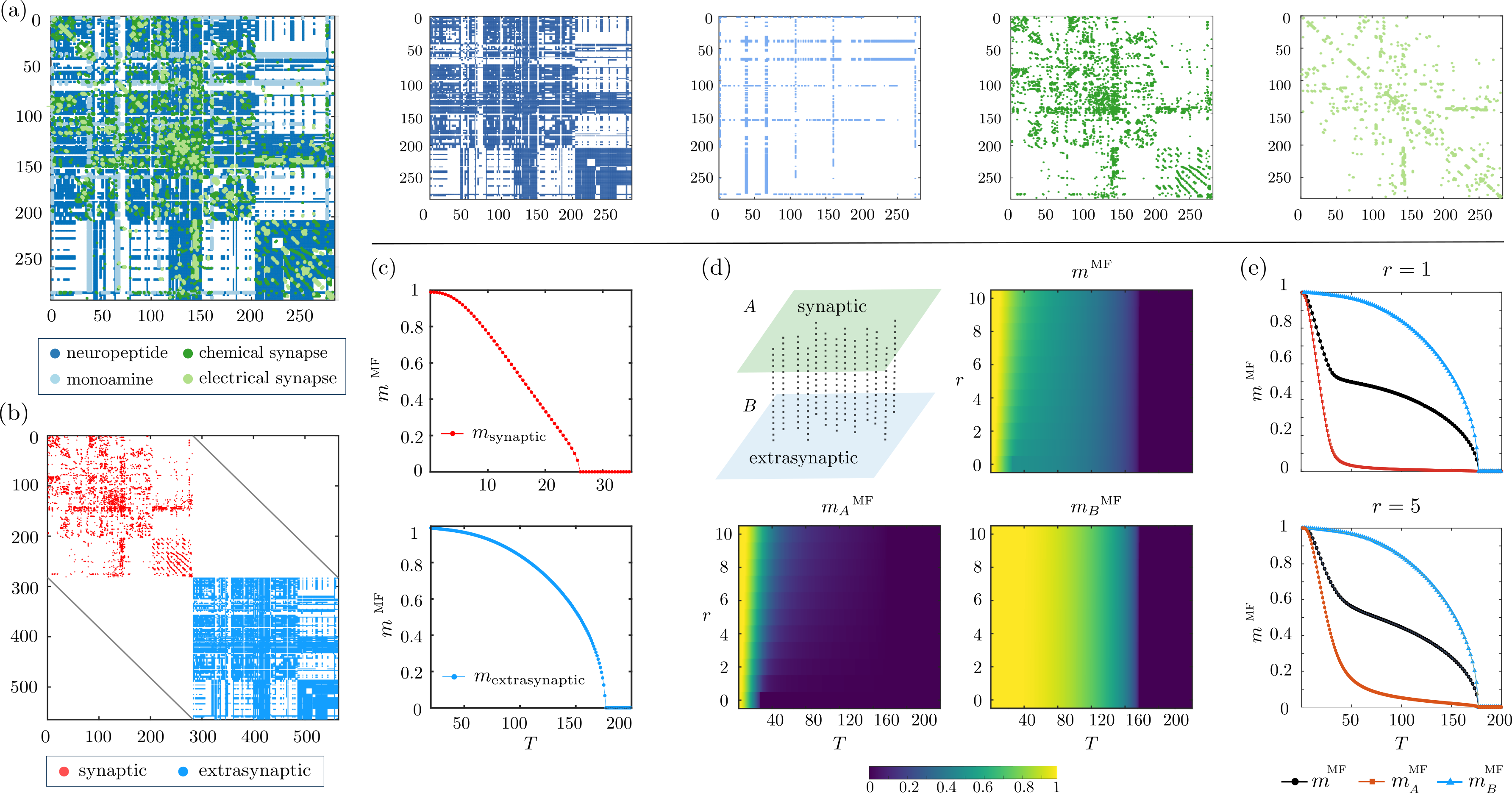}
    \caption{We examine Ising dynamics on a two-layer multiplex network of distinct signaling modes between neurons in the nematode \emph{C. elegans} using the mean-field approximation in Sec.~\ref{sec:mean_field_approx}. (a) Adjacency matrices corresponding to distinct signaling modes between neurons in the somatic nervous system of \emph{C. elegans}. The leftmost plot overlays the adjacency matrices of all four signaling modes. The right four plots show the adjacency matrices of each individual signaling mode. In all plots, the horizontal and vertical axes indicate neuron indices. (b) The supra-adjacency matrix of an associated two-layer multiplex network. The first layer is an aggregate network of chemical and electrical synapses (i.e., synaptic signaling), and the second layer is an aggregate network of neuropeptide and monoamine signaling (i.e., extrasynaptic signaling). (c) Mean-field magnetization as a function of temperature $T$ for the aggregate networks of (top) synaptic signaling and (bottom) extrasynaptic signaling. (d) We show (top left) a schematic illustration of a multiplex network of synaptic and extrasynaptic signaling, (top right) a heat map of the mean-field magnetization of the overall multiplex network, (bottom left) a heat map of the mean-field magnetization of the synaptic layer, and (bottom right) a heat map of the mean-field magnetization of the extrasynaptic layer. In each heat map, the horizontal axis indicates the temperature $T$ and the vertical indicates the interlayer coupling strength $r$. (e) Mean-field magnetizations of the overall multiplex network, the synaptic layer, and the extrasynaptic layer as functions of temperature for interlayer correlation strengths of (top) $r = 1$ and (bottom) $r = 5$. }
\label{fig:c_elegans}
\end{figure*}

The nematode \emph{C. elegans}, which has 302 neurons, is a relatively simple organism. Its anatomical connectome, which has both chemical synapses and electrical synapses, has been mapped using electron microscopy \cite{white1986structure, varshney2011structural}. As with all other organisms, the neural activity in \emph{C. elegans} arises from an interplay between distinct modes of signaling. Moreover, in addition to synaptic transmission, recent studies of \emph{C. elegans} have highlighted the role of extrasynaptic signaling that arises via the diffusion of neuropeptides and monoamines~\cite{bentley2016multilayer, randi2023neural, ripoll2023neuropeptidergic}. Accordingly, to develop a more complete picture of the nervous system, it is important to account for multiple signaling modes, examine the structures and interdependencies of these modes, and assess their influence on neuronal dynamics.

One can represent the network structure of neurons and their multimodal signaling interactions as a multilayer network \cite{bentley2016multilayer}. Investigations of neuronal systems have focused primarily on synaptic networks (i.e., structural connectomes)~\cite{varshney2011structural, towlson2013rich}, but extrasynaptic signaling can also significantly influence processes on connectomes~\cite{marder2012neuromodulation, bargmann2012beyond, randi2023neural}.

As an illustrative example to explore how incorporating multiple signaling networks can affect qualitative dynamics on neuronal networks, we study Ising dynamics on a multiplex connectome. We obtain the chemical and electrical synapse networks from Ref.~\cite{varshney2011structural}, and we obtain the extrasynaptic signaling networks from Ref.~\cite{ripoll2023neuropeptidergic}. We restrict our study of the \emph{C. elegans} multimodal signaling network to the somatic nervous system, which consists of $N = 281$ neurons. To visualize the distinct network structures of each signaling mode, we show the unweighted adjacency matrices with aligned neuron indices in Fig.~\ref{fig:c_elegans}(a). We examine how accounting for the extrasynaptic network (in the form of signaling through monoamines and neuropeptides) can impact the Ising dynamics. 

To obtain a multiplex network, we first construct an aggregate network of chemical and electrical synapses that we call the ``synaptic network" and an aggregate network of neuropeptide and monoamine signaling that we call the ``extrasynaptic network". For simplicity, we consider undirected and unweighted network architectures on each layer. The synaptic network has a mean degree of $k \approx 8$, and the extrasynaptic network is much denser, with a mean degree of $k \approx 79$. We assemble a two-layer multiplex network with the synaptic network as layer $A$, the extrasynaptic network as layer $B$, and interlayer edges that capture interactions between corresponding state nodes of the two signaling modes. For simplicity, we assume that all interlayer edges have the same weight $r$. In Fig. \ref{fig:c_elegans}(b), we show the supra-adjacency matrix of this two-layer multiplex network.

We first examine Ising dynamics on the synaptic and extrasynaptic networks individually using the mean-field approximation in Sec.~\ref{sec:mean_field_approx}. In Fig.~\ref{fig:c_elegans}(c), we show the mean-field magnetization of each network as a function of temperature. The mean-field critical temperature of the synaptic network is $T_c \approx 26$, and the mean-field critical temperature of the extrasynaptic network is $T_c \approx 177$. We then examine mean-field Ising dynamics on the two-layer multiplex network as a function of the temperature $T$ and the interlayer coupling strength $r$. In Fig. \ref{fig:c_elegans}(d), we show the resulting heat maps of the overall and single-layer mean-field magnetizations. 

We observe that increasing the interlayer coupling strength $r$ influences the overall magnetization of the multiplex network. This increase arises primarily through changes in the magnetization of the synaptic layer; the value of $r$ has no noticeable effect on the extrasynaptic layer for the examined values of $r$. However, although the magnetization of the extrasynaptic layer does not change noticeably, the presence of this layer causes the magnetization of the synaptic layer to approach $0$ at a higher temperature than it would in isolation. This observation is consistent with how extrasynaptic signaling is often described in neuroscience~\cite{marder2012neuromodulation}. Extrasynaptic signaling typically does not generate large-scale activity on its own. Instead, it shapes circuit dynamics by altering the conditions in which synaptic networks transition between different states~\cite{marder2012neuromodulation}.

To illustrate more clearly how the overall and single-layer magnetizations depend on temperature $T$ and interlayer coupling strength $r$, we plot [see Fig.~\ref{fig:c_elegans}(e)] selected rows of the heat maps from Fig.~\ref{fig:c_elegans}(d) to show the dependence of the magnetizations on $T$ for $r = 1$ and $r = 5$.


\subsection{Multi-university Facebook network}\label{sec:fb_nets}

As a second example of a real-world multilayer network, we examine Ising dynamics on a multi-university Facebook network of user accounts from University of Michigan and Michigan State University~\cite{melnik2014dynamics,traud2012}. This network includes both intra-university and inter-university Facebook ``friendships" between accounts.

\begin{figure*}
    \centering
    \includegraphics[width=\linewidth]{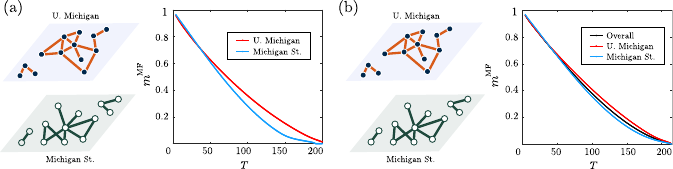}
    \caption{We examine Ising dynamics on a multi-university Facebook friendship network of user accounts from University of Michigan and Michigan State University~\cite{melnik2014dynamics,traud2012} using the mean-field approximation in Sec. \ref{sec:mean_field_approx}. (a) In the left panel, we show a schematic illustration of the two single-university networks. In the right panel, we show the mean-field magnetization of the largest connected component of each university's Facebook network as a function of temperature. (b) In the left panel, we show a schematic illustration of the two-layer Facebook network. The dashed lines indicate inter-university friendships. In the right panel, we show the mean-field magnetizations of the overall multilayer network and of the largest connected component of each layer. }
    \label{fig:FB_networks}
\end{figure*}

We first study Ising dynamics on the largest connected component (LCC) of each individual Facebook network. The LCC of the University of Michigan network has $N = 30106$ nodes and a mean degree of $k \approx 78$, and the LCC of the Michigan State University network has $N = 32361$ nodes and a mean degree of $k \approx 69$. We compute the mean-field magnetization of each Facebook network as a function of temperature using Eq.~\eqref{eq:self_consistency1}, and we show our results in Fig.~\ref{fig:FB_networks}(a). For both networks, we observe that the magnetization decreases gradually and does not appear to have sharp transitions. The temperatures at which the networks have a mean-field magnetization of $m^{\text{MF}} = 0.1$ are $T \approx 165.5$ for the University of Michigan network and $T \approx 138$ for the Michigan State University network, respectively.

We then consider a multilayer network with interlayer edges that correspond to Facebook friendships between individuals at different universities. This multilayer network thus has a different structure from a multiplex network~\cite{kivela2014multilayer}. The LCC of this multilayer network has $N = 62770$ state nodes. We analyze Ising dynamics on the LCC of this network and track the mean-field magnetization of the overall network and of the LCCs of each individual layer. We show our results in Fig. \ref{fig:FB_networks}(b). In comparison to Fig.~\ref{fig:FB_networks}(a), we observe that the magnetization of each layer’s LCC is larger in the multilayer network than in the isolated networks.  The effect is particularly pronounced for the Michigan State University network, which is the sparser of the two networks. This observation is also consistent with our previous observations for coupled networks with different mean degrees (see Sec.~\ref{sec:ER_diff}). The temperatures at which the mean-field magnetization is $m^{\text{MF}} = 0.1$ are approximately $T \approx 168$ for the overall multilayer network, $T \approx 175.7$ for the University of Michigan layer, and $T \approx 160$ for the Michigan State University layer. 
%


\section{Conclusions and discussion}\label{sec:discussion}

When studying a dynamical process on a multilayer network, it is important to consider how both layer structure and cross-layer heterogeneities impact the qualitative and quantitative dynamics. To illustrate this point, we studied the dynamics of a multiplex ferromagnetic Ising model in which nodes have distinct spins on different layers. We emphasized the effects of cross-layer heterogeneities, which are common in real-world systems \cite{kivela2014multilayer,aleta2019multilayer,de2023more}, and we explored how layer structure and interlayer coupling influence the multilayer dynamics. Using both synthetic and empirical networks, we demonstrated that cross-layer heterogeneity can significantly influence Ising dynamics on multilayer networks.

We investigated Ising dynamics on a variety of synthetic and real-world two-layer networks in which all interlayer edges have the same weight, which encodes the interlayer coupling strength. We examined the influence of interlayer coupling strength and temperature on the steady-state behavior of the model using Monte Carlo (MC) simulations and a mean-field approximation. We observed that increasing the interlayer coupling strength increases the temperature at which the magnetization of the network approaches $0$. In our two-layer multiplex networks with layers of different densities, this phenomenon was particularly noticeable in the sparser layer. The coupling of the sparser layer to a denser layer causes the former to retain a nonzero magnetization above its original critical temperature.

Using a mean-field approximation, we examined the stability of an anti-aligned solution, in which the two layers have magnetizations of opposite sign. For the synthetic networks that we examined, we observed that increasing the interlayer coupling strength narrows the temperature range with metastable anti-aligned solutions. Additionally, we observed that the anti-aligned solution loses metastability at a weaker interlayer coupling and is metastable for a smaller temperature range when the layers have different mean degrees than when they have the same mean degree. By incorporating core--periphery (CP) structure into one layer of two-layer multiplex networks, we observed that interlayer edges that involve peripheral nodes can exert more influence on dynamics than interlayer edges that involve only core nodes. This observation suggests that peripheral nodes can play important roles in helping to shape system-wide behavior. Finally, we examined Ising dynamics on two empirical two-layer networks---a neural-signaling network and a multi-university Facebook network---with heterogeneous layers. 

There are many avenues for further study. Naturally, one can consider a variety of different network architectures and dynamical processes. In particular, we expect that it will be very valuable to systematically examine the effects of layer heterogeneities---both within and across layers---on a range of dynamical processes. Insights from such investigations may also inform the design of targeted intervention strategies to influence dynamical processes on multilayer networks. Additionally, expanding analysis from theoretically-motivated settings to more realistic scenarios can yield important insights into multilayer biological and social networks. In the next several paragraphs, we expand on these ideas.

We only considered a few types of synthetic networks, and it is valuable to study multilayer Ising dynamics on networks with community structure, other mesoscale features, and cross-layer structural correlations. In Sec.~\ref{sec:CP_ER}, we examined a synthetic two-layer network in which one layer had CP structure and the other layer was an Erd\H{o}s--R\'{e}nyi (ER) network. A natural extension of this example is to consider multilayer CP networks with cross-layer heterogeneities in which nodes belong to the core set of one layer but to the periphery set of another layer. Because cross-layer heterogeneities are important for Ising dynamics (and undoubtedly also for other dynamical processes), we particularly encourage investigations that further explore how heterogeneity in interlayer couplings affect dynamical processes on multilayer networks. It is also worth studying how metastable states, such as ones in which different layers or different communities have opposite magnetizations, can arise from network architectures with structural correlations. 

Naturally, there are many possible extensions of Ising dynamics, including ones that expand beyond binary spin values~\cite{castellano2009statistical} and others that incorporate polyadic (i.e., ``higher-order") interactions~\cite{robiglio2024higher}. One particularly interesting direction is to incorporate external fields, such as a global field that depends on the state of a system~\cite{devauchelle2024dislike} or fields that are different for different nodes~\cite{lynn2017statistical}. In exploring Ising dynamics and its extensions, a key consideration is how external interventions can alter network dynamics to promote or destabilize particular states. More generally, such external interventions can act on nodes, edges, or other network structures. These avenues complement existing work on the control of dynamical processes on multilayer networks \cite{srivastava2021structural,menichetti2016control,posfai2016controllability}.

Additionally, although we examined Ising dynamics, we expect that the phenomena that we explored are relevant more generally for dynamical processes on multilayer networks. The influence of cross-layer structural heterogeneities on dynamical processes has been explored predominantly in theoretically-minded settings \cite{de2016physics,de2023more}, and it is important to conduct systematic investigations of the impact of such heterogeneities in more realistic scenarios. In the next two paragraphs, we discuss one application in neuroscience and one application in the social sciences.

In neuroscience, the influence of the extrasynaptic signaling network on the multiplex Ising dynamics in \emph{C. elegans} highlights the importance of investigating models that couple synaptic neuronal activity with diffusion processes through extrasynaptic pathways. Related toy models that consider distinct, coupled dynamical processes on multiplex networks---such as coupled oscillators (to model neural dynamics) with random-walk dynamics (to model blood flow)~\cite{nicosia2017}---may provide starting points to develop more realistic multilayer models. More generally, one can use multilayer frameworks to capture information from multimodal, multiscale, and spatiotemporal data from modern neuroscience experiments \cite{vaiana2020multilayer,kulkarni2025toward}, and it is particularly important to develop a thorough understanding of how intralayer and interlayer structural heterogeneities affect associated dynamical processes. 

In the social sciences, multilayer network models in which the same individual can have distinct opinions (or distinct values of other variables) in different layers are relevant for opinion dynamics. One can use such settings to study opinion dynamics on interdependent topics \cite{li2025bounded,xiong2017analysis} and to distinguish between intrinsic and expressed (or perceived) opinions \cite{goel2010real, zimmaro2025meta}. Models that account for differences in how individuals express their opinions can be particularly interesting for networks with community structure or CP structure, as individuals may express their beliefs differently in different social settings.

For example, individuals may share information to different extents when they navigate different communities \cite{baek2023perceived} or they may adjust their expressed views based on their perception of what is the majority opinion \cite{gaisbauer2020dynamics}. There have not been many systematic investigations of such distinctions in models of opinion dynamics, and it is worthwhile to examine opinion models with such features on structurally heterogeneous social networks. Such investigations may lead to interesting insights into how network heterogeneities combine with how individuals navigate social networks to shape opinion dynamics.

\section*{Acknowledgements}

We thank Deepak Dhar, Aanjaneya Kumar, and Aaron Winn for helpful discussions. S.S.K and D.S.B acknowledge support from the Army Research Office MURI program (through grant number W911NF2410228). C.W.L. acknowledges support from the National Institutes of Health (through grant number R35GM160188).


\section*{Code and data availability}

Our simulation code and a subset of our data are available at \href{https://github.com/SumanSKulkarni/Ising-Multilayer}{https://github.com/SumanSKulkarni/Ising-Multilayer}.


\appendix


\section{Mean-field approximation}\label{sec:meanfield_derivation}

The Hamiltonian of the ferromagnetic Ising model with no external field is
\begin{equation}
    \mathcal{H}(\mathbf{s}) = - \sum_{i,j} J_{ij} s_i s_j \,,
\end{equation}
where $J_{ij} > 0$ is the strength of the interaction between node $i$ and node $j$. The probability distribution over node configurations is given by the Boltzmann probability distribution
\begin{equation}
    P(\mathbf{s}) = \frac{e^{-{ \mathcal{H}(\mathbf{s})}/T}}{\mathcal{Z}}\,,
\end{equation}
where 
\begin{equation}
    \mathcal{Z} = \sum_{\mathbf{s}} e^{-{ \mathcal{H(\mathbf{s})}/T}}
\end{equation}
is the partition function.

To obtain a mean-field approximation of the Gibbs free energy, we consider a trial probability function in the product form
\begin{equation} 
    \textrm{Prob}(\{s_i\}) = \prod_{i} \left[\frac{1 + m_i}{2} \, \delta(s_i - 1) + \frac{1 - m_i}{2} \, \delta(s_i + 1) \right]\,.
    \label{eq:trial_prob}
\end{equation}
Using the probability distribution \eqref{eq:trial_prob}, we solve self-consistently for the expected value of the magnetization $m_i$ of node $i$ and thereby obtain the equation
\begin{equation}
    m_i = \tanh \left( \frac{\sum_{j} J_{ij} m_j}{T}\right) 
    \label{eq:self_consistency2}
\end{equation}
for each node $i$. Under this mean-field approximation, the Gibbs free energy is
\begin{widetext}
\begin{equation}
    G_{\textrm{MF}} = - \sum_{i,j} J_{ij} m_i m_j + T \left\{ \sum_{i} \left[ \frac{1 + m_i}{2} \ln \left( \frac{1 + m_i}{2} \right) + \frac{1 - m_i}{2} \ln \left( \frac{1 - m_i}{2} \right) \right] \right\} \,,
    \label{eq:free_energy2}
\end{equation}
\end{widetext}
where the first term is a Hamiltonian term and the second term is an entropy term. 
We examine the stability of solutions using the Gibbs free energy \eqref{eq:free_energy2}. The first derivative of the free energy with respect to the magnetization $m_i$ is
\begin{equation}
    \frac{\partial G}{\partial m_i} = -\sum_j J_{ij}m_j + T \tanh^{-1}(m_i) \,.
\end{equation}
By setting these derivatives to 0, we again obtain the same self-consistency equations \eqref{eq:self_consistency2}. We also compute the second-derivative matrix (i.e., the Hessian matrix) of the free energy. Its entries are
\begin{equation}
    H_{ij} = \frac{\partial^{2} G}{\partial m_j m_i} = - J_{ij} + \frac{T}{1 - m_i^2} \, \delta_{ij} \,.
    \label{eq:hessian_ij}
\end{equation} 
In matrix form, the Hessian is
\begin{equation}
    H(G) = - J + T \, \textrm{Diag} \left[ \frac{1}{1 - m_i^2}\right] \,.
    \label{eq:hessian}
\end{equation}

There are many possible solutions of the self-consistency equations \eqref{eq:self_consistency2}. These solutions can be local minima, local maxima, or saddle points of the Gibbs free energy \eqref{eq:free_energy2}. A solution of the self-consistency equations is a local minimum of the Gibbs free energy $G_{\textrm{MF}}$ if all of the eigenvalues of the Hessian are positive, so we are most interested in this situation. Because we consider real and symmetric $J$, the Hessian is also real and symmetric and thus has real eigenvalues.

Equilibrium solutions of Eq.~\eqref{eq:self_consistency2} correspond to global minima of the mean-field Gibbs free energy \eqref{eq:free_energy2}. For ferromagnetic coupling matrices (i.e., $J_{ij} > 0$), the Gibbs free energy is symmetric under a global spin flip, so any solution $\{m_i\}$ has a counterpart $\{-m_i\}$ with the same free energy. At low temperatures, the Gibbs free energy \eqref{eq:free_energy2} is dominated by the Hamiltonian term. Consequently, we expect all nodes to have spins of the same sign in equilibrium solutions. At high temperatures, the Gibbs free energy \eqref{eq:free_energy2} is dominated by the entropy term. In particular, when $T > \lambda_{\textrm{max}}(J)$ (where $\lambda_{\textrm{max}}(J)$ is the largest eigenvalue of J), the Gibbs free energy is strictly convex and the equilibrium solution is unique. By symmetry, we expect that $m_i = 0$ for all $i$ in the equilibrium solution. Depending on the network structure and model parameters, there may exist solutions of Eq.~\eqref{eq:self_consistency2} that are local minima of Eq.~\eqref{eq:free_energy2}. These solutions are ``metastable" states. A system that is in or sufficiently near such a metastable state can remain trapped in that local minimum. 

\bibliography{ising_multilayerMAIN-v08.bib}

\end{document}